# TWO- FINGER KEYBOARD LAYOUT PROBLEM: AN APPLICATION ON TURKISH LANGUAGE

Kürşad Ağpak[a], Hüseyin Karateke[b], Süleyman Mete[a]

[a]Gaziantep University, Industrial Engineering Department

[b]Gazi University, Industrial Engineering Department

**Abstract**

Smart phone and tablet usage has sharply increased for the last decade. While entering test on these devices, virtual keyboards are generally used instead of conventional hardware keyboards. In this study, a new problem which is two-finger keyboard layout problem and solution approach is presented for increasing user test entrance performance, especially on virtual keyboards. Defined two-finger keyboard layout problem is modeled as Quadratic Assignment Problem. Because of combinatorial structure of the problem a genetic algorithm is developed. Its result is given to mathematical model as initial solution for finding better solutions with mathematical model. Proposed approach is applied on Turkish language. The new two finger keyboard layout for Turkish language is compared with F and QWERTY keyboard layouts based on certain performance measurement techniques.

**Keywords**: Two-finger keyboard layout problem, quadratic assignment problem, genetic algorithm, virtual keyboard.

## 1. Introduction

Continuing development of smart phones and tablets has been making communication easier and faster recently. Therefore, usage of smart phone and tablet is increasing rapidly day by day. Generally, virtual keyboards are used to enter text document on tablets and smart phones. Literature search show that there are several type of keyboard designs for different languages such as Opti (Mac Kenzie & Zhang, 1999), Metropolis (Zhai et al., 2000), Ylarof (Li et al., 2006) and Fitaly (Fitaly, 2013) for English, AZERTY for French and F for Turkish. While developing keyboard design, certain criteria are considered, e.g. ergonomic factors or writing performance. Today keyboards, especially virtual ones, are used not only certain type of profession as writers or journalists but also used by adults to primary school students lots of kind of people which are using smartphone or tablet for social sites or blogs on web, chatting each other etc. But we can say that some users are novice and use less than 10 fingers while writing. And lots of them can be use only two finger on virtual keyboard due to dimensional restrictions of smartphones and tablets. Therefore in this study new type of keyboard arrangement problem which is two-finger keyboard layout problem is defined, and methodology presented to solve the problem. Purpose of new design problem is ensuring to enter text fast and with less finger movements. Problem has been modelled as a type of QAP. Nevertheless the optimal solution is difficult by using exact algorithms because of complexity of QAP which is NP-hard. Therefore, in literature several heuristic methods such as taboo search algorithm, genetic algorithm, simulated annealing, particle swarm optimization method and ant colony algorithm, are developed to find solution QAPs (Burkard, 2013). In this study, a simple genetic algorithm (GA) is developed to solve the problem. Defined problem is solved for Turkish language and found keyboard is compared with F and QWERTY keyboards.

The paper is arranged as follows; in the second section two finger keyboard layout problem is defined and QAP model is presented. Two finger stylus keyboard (Ç) for Turkish language is designed in the third section. Proposed Ç keyboard, F keyboard and QWERTY keyboard are compared with each other in the fourth section. The last section is conclusion and future studies.

## 2. Problem Definition

For the two finger keyboard layout problem, standard longitudinal shape keyboard is considered as two part, left and right. When entering text, it is assumed that users use only one the left and right fingers on the left and right parts, respectively (see Figure 1). Objective of problem is minimization of sum of finger movements on two keyboard parts.

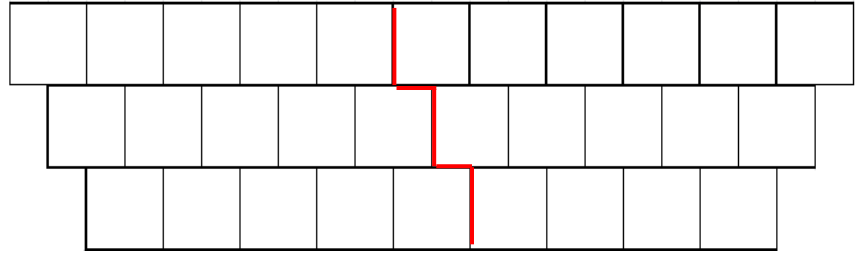

Figure 1.Traditional Longitudinal Keyboard Layout

The problem is modeled as a type of QAP similar to single-finger keyboard problem. Assumptions of model are given below.
- ✓ Keys for assigning to locations are known and fixed.
- ✓ Keyboard layout and locations on keyboard to assign keys are fixed and known.
- ✓ Euclidean distance is taken for distance between locations.
- ✓ Key digraph matrix is deterministically given for language.
- ✓ Keyboard is separated to two as left and right part.
- ✓ While entering text, keyboard user uses left and right hand for keys on left and right parts respectively.

Under these assumptions, the problem is to assign keys to locations such that total movements of fingers will be minimized. Developed QAP model is given below.

$$min \sum_{m=1}^{2} \sum_{i=1}^{n} \sum_{j=1}^{n} \sum_{k=1}^{n} \sum_{l=1}^{n} C_{ik} * D_{jl} * XY_{imj} * XY_{kml} \qquad (1)$$

Subject to

$$\sum_{m=1}^{2} Y_{im} = 1 \qquad for\ all\ i \qquad (2)$$

$$\sum_{j=1}^{n} XY_{imj} - Y_{im} = 0 \qquad for\ all\ i, m \qquad (3)$$

$$\sum_{i=1}^{n} XY_{imj} = 1 \qquad \text{for all } j, m \qquad (4)$$

$$XY_{imj}, Y_{im} \in \{0,1\} \qquad (5)$$

There are two variables, $Y_{im}$ and $XY_{imj}$. $Y_{im}$ takes 1 if key i is assigned to part m, otherwise takes 0. In a similar way, $XY_{imj}$ is takes 1 if key i is assigned to location j on part m. Each value on key digraph matrix, $C_{ik}$, is shows that number of times of key k comes consecutively after key i in selected texts. $D_{jl}$ is Euclidean distance between location j and l. Objective function minimizes sum of all movements based on $C_{ik}$ and $D_{jl}$ matrixes. Constraint 2 ensures that each letter must be assigned to one part, left or right. If $Y_{im}$ takes 1, key i assigned to one location on part m with Constraint 3. Constraint 4 provides that one key must be assigned to each location on each part.. Last constraint (5) is the the binary constraints.

### 3. Designing Two Finger Keyboard Layout for Turkish Language

Special keyboard design for Turkish language has been firstly designed by İhsan Sıtkı Yener in 1955 which is called as F keyboard. F keyboard is designed for ten fingers and has been widely used in Turkey with state support. On the other hand, stylus keyboard for Turkish language has been firstly presented by Uşşaklı (2004). Then Soydal (2010) presented another design. Two new designs are developed for single finger. In this study, a new two finger keyboard layout (TFKL) for Turkish is designed for novice users to type fast with two fingers. There are five steps to design TFKL for any language. These steps are presented in Figure 2.

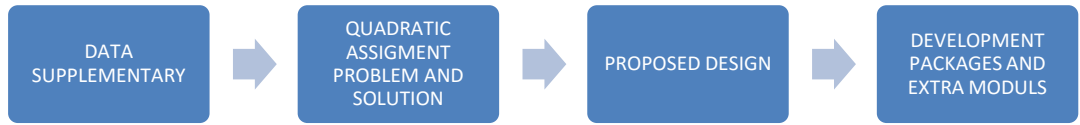

Figure 2. Steps of Designing TFKL for Turkish Language

**Step 1. Data Supplementary**
First step of designing TFKL is data supplementary. Data supplementary has two parts. One part of data supplementary is frequency number of characters. The other part of data supplementary is distance between keys. These two sub-steps are explained below in detail.
**Flow matrix:** Key digraph matrix, $C_{ik}$, is formed in this step. Turkish language have 29 letters, they are shown below. Letters out of these (*q, w, x*), space bar and punctuation characters are ignored.

*Set of letters for Turkish language= {a, b, c, ç, d, e, f, g, ğ, h, ı, i, j, k, l, m, n, o, ö, p, r, s, ş, t, u, ü, v, y, z}*

Base texts for generating digraph are selected from Turkish web blogs and several recent published books. The result of this step, 29x29 dimensions asymmetric matrix for Turkish language is established and given in Appendix A.

**Distance matrix:** Distance matrix, $D_{jl}$, is calculated by using Euclidian distance. In this study, the geometrical shape of keyboard is standard longitudinal shape. Two zones with 15 and 14 key locations are determined for possible assignment of 29 Turkish language letters as Figure 3. Distance between center points of two horizontal adjacent keys is one unit. Vertical distance of two rows is also one unit. For example, distance between key 1 and key 2 one unit, key 1 and key 6 is $\sqrt{1,25}$. Full distance matrix is given in Appendix B.

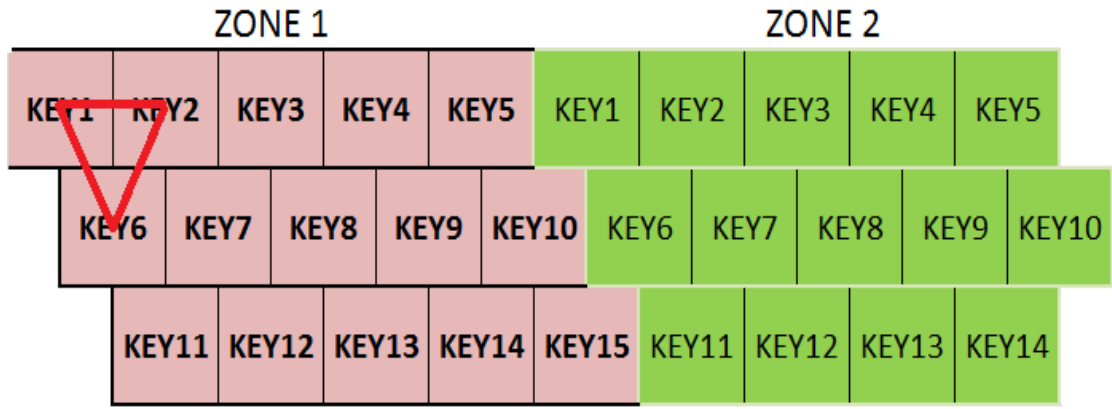

Figure 3. New Proposed Keyboard Layout.

**Step 2. Quadratic Assignment Problem and Solution**

There are two steps to solve TFKL problem. These are genetic algorithm (GA) methodology and exact algorithm. Basic GA structure is used and coded in MATLAB, steps of algorithm and selected parameters can be found in Appendix C. Mathematical model is coded in General Algebraic Modeling System (GAMS). Firstly, TFKL problem is solved with GA. In second stage, best solution of GA is given to the mathematical model as initial solution to find better solution. Mathematical model is solved with GAMS-Cplex 12.3 which is run two week. But solution is not improved due to NP-hard structure of problem. So, GA results are taken as a solution for the design.

**Step 3. Proposed design for two finger stylus keyboard layout**

The best arrangement of TFKL problem for Turkish language is given in Figure 4. The name of proposed model is Ç keyboard. This comes from its first character. All vowel letters are located in right side. The reason of this can be that vowel and consonant letters generally follows each other in Turkish language.

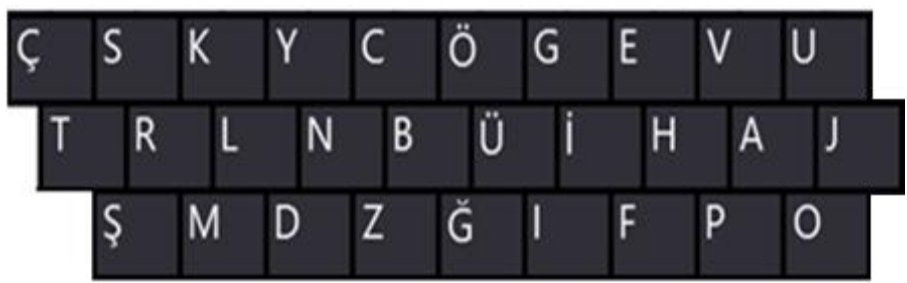

Figure 4. Proposed Keyboard Layout (Ç keyboard)

**Step 4. Packages and Extra Module**

In this step, Microsoft and Android applications of Ç keyboard are developed to apply usability test for participants. There are two screenshots for android application for Ç keyboard. These are vertical and horizontal screenshots in Figure 6 and Figure 7, respectively. Microsoft application screenshot is available in Figure 5.

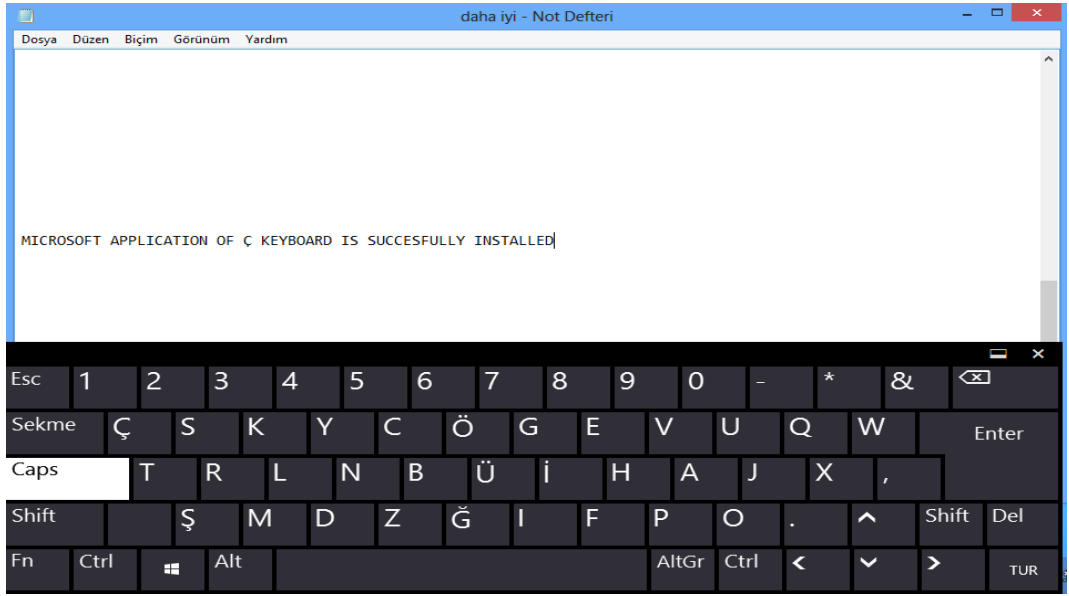

Figure 5. Screenshot of Microsoft Application of Ç Keyboard

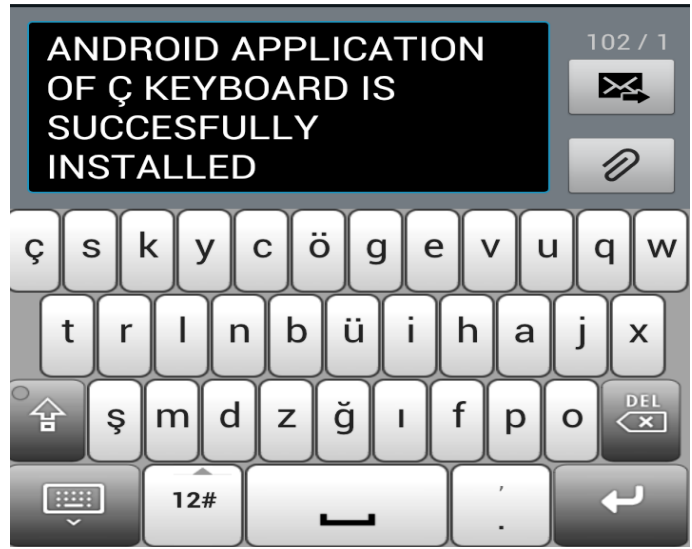

Figure 6. Vertical layout on Android application of Ç keyboard

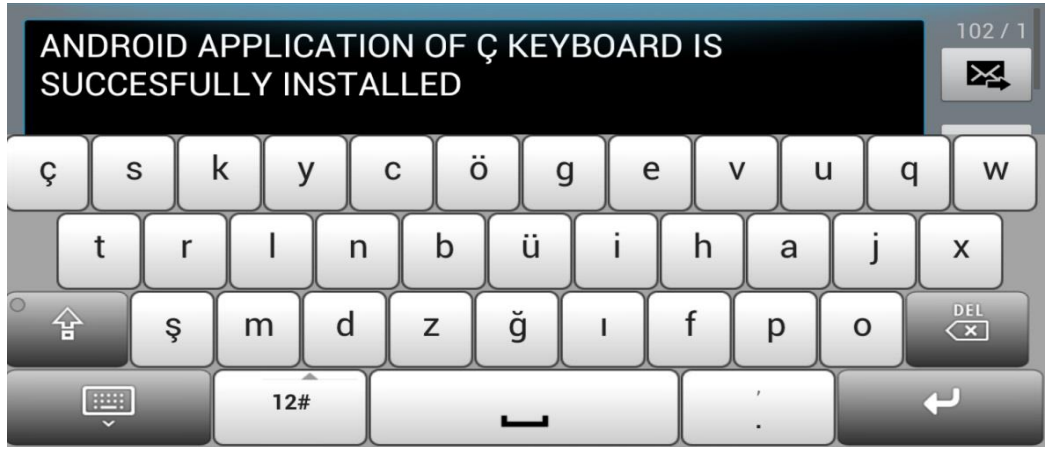

Figure 7. Horizontal layout on Android application of Ç Keyboard

## 4. Discussions and Results

Objective function values and usability analysis are used to compare Q, F and Ç keyboards. Values of objective function of mathematical model are calculated using Equation (1) for Ç, F and Q keyboards, which are shown in Table 1.

Table 1. Objective Function Values of Keyboards

|  | Ç Keyboard | F Keyboard | Q Keyboard |
| --- | --- | --- | --- |
| Objective Values | 202,601 | 310,723 | 825,800 |

Ç keyboard is better than other two keyboards according to objective function values. Q keyboard is the worst because its development process does not include any issue about Turkish language.

Usability is defined as: "the extent to which a product can be used by specified users to achieve specified goals with effectiveness; the extent to which the intended goals of use are achieved, efficiency; the resources that have to be expended to achieve the intended goals and satisfaction; the extent to which the user finds the use of the product acceptable, in a specified context of use" (ISO 9241-11,1994). There are three usability evaluation methods in literature. One of these methods is usability test. The usability test requires sample participant to work on typical tasks using the system. The evaluators use the results to see how the user interface supports the users to do their tasks. In this section, Ç, F and Q keyboards are compared based on efficiency and effectiveness using usability test. The layouts are depicted to the subjects on the test software for Android system. It is said these subjects that "both accuracy and typing speed is important for this experiment". Also, it is clearly implied to them that the time measurements will not be shown to the other subject, in order to prevent a competition between the subjects. Each participant was given oral instructions explaining the task and the goal of the experiment. They were asked specifically to aim for both entry speed and accuracy. The instructions also stated that if they made an error, that try will be repeated. In other words, a session is repeated if the text typed in that session has missing, extra or different characters when compared with the test material. Hence, after each session, error rate should be zero. Three tasks are applied to practical usability test. First task is to write specific text which has 31 words and 212 characters. Second task is to enter a tweet which has 20 words and 140 characters. Third task is to write e-mail text which has 29 words and 201 characters. Results are shown in Table 2.

Table 2. Text entry seconds for Ç, F and QWERTY(Q) keyboard for Task 1, Task 2 and Task3

| Participants | Task 1 | | | Task 2 | | | Task 3 | | |
| --- | --- | --- | --- | --- | --- | --- | --- | --- | --- |
| | Q | F | Ç | Q | F | Ç | Q | F | Ç |
| Participant 1 | 169 | 529 | 402 | 148 | 288 | 201 | 180 | 596 | 459 |
| Participant 2 | 176 | 317 | 311 | 164 | 471 | 388 | 189 | 437 | 378 |
| Participant 3 | 150 | 446 | 283 | 158 | 428 | 372 | 169 | 574 | 342 |
| Participant 4 | 210 | 477 | 384 | 178 | 443 | 309 | 243 | 547 | 442 |
| Participant 5 | 162 | 510 | 420 | 152 | 468 | 361 | 175 | 597 | 463 |
| Participant 6 | 292 | 403 | 371 | 236 | 378 | 327 | 261 | 422 | 396 |
| Participant 7 | 115 | 250 | 205 | 97 | 231 | 140 | 126 | 376 | 260 |
| Participant 8 | 152 | 398 | 267 | 141 | 458 | 346 | 168 | 521 | 351 |
| Participant 9 | 171 | 365 | 281 | 198 | 434 | 338 | 188 | 466 | 361 |
| **AVERAGE** | **177,444** | **410,555** | **324,888** | **163,555** | **399,888** | **309,111** | **188,777** | **504** | **383,555** |

According to Table 2, Q keyboard is best keyboard for all tasks because it is the most known and used keyboard by the users. But when the two unfamiliar keyboards are compared, we see that new design is significantly better than other for three tasks.

## 5. Conclusion and Future Studies

In this study, two finger keyboard layout problem is defined and mathematically modelled as QAP. Defined problem is solved for Turkish language. The proposed design, Ç keyboard layout, is compared with well-known Q keyboard and F keyboard which is the first keyboard designed for Turkish language, according to QAP model objective value and usability test. Although Ç keyboard has better objective value than two keyboards, it is usability test results are worse than Q keyboard. Because finding inexperienced Q keyboard user is pretty hard. In future studies TFKL problem can be solved for other languages. Also new techniques can be developed to solve the problem efficiently.

# APPENDIX A. CHARACTER TRANSITION FREQUENCE (CTF) OR FLOW MATRICE for TURKISH LANGUAGE

Character Transition Frequence (CTF) of Flow Matrice for Turkish Language

|   | A | B | C | Ç | D | E | F | G | Ğ | H | I | İ | J | K | L | M | N | O | Ö | P | R | S | Ş | T | U | Ü | V | Y | Z |
|---|---|---|---|---|---|---|---|---|---|---|---|---|---|---|---|---|---|---|---|---|---|---|---|---|---|---|---|---|---|
| A | 459 | 2728 | 1403 | 914 | 4589 | 201 | 2856 | 1046 | 305 | 1711 | 2591 | 4 | 580 | 289 | 7160 | 8505 | 4515 | 15490 | 98 | 0 | 2142 | 16610 | 4276 | 2972 | 4766 | 133 | 1425 | 5414 | 2239 |
| B | 4144 | 50 | 5 | 0 | 40 | 2856 | 2 | 2 | 0 | 0 | 5 | 10564 | 0 | 5 | 18 | 194 | 0 | 979 | 510 | 1 | 226 | 4 | 0 | 0 | 8 | 4139 | 700 | 0 | 29 |
| C | 2722 | 10 | 29 | 0 | 9 | 3307 | 0 | 0 | 1 | 0 | 426 | 947 | 0 | 0 | 264 | 112 | 14 | 409 | 2 | 1 | 210 | 3 | 0 | 0 | 49 | 1067 | 133 | 4 | 3 |
| Ç | 1173 | 60 | 0 | 2 | 0 | 1706 | 0 | 0 | 0 | 0 | 1069 | 2384 | 0 | 0 | 58 | 294 | 303 | 3 | 73 | 0 | 0 | 18 | 0 | 0 | 161 | 93 | 311 | 0 | 0 |
| D | 11666 | 5 | 5 | 0 | 268 | 11085 | 0 | 6 | 16 | 0 | 20 | 2676 | 0 | 0 | 0 | 78 | 9 | 1732 | 523 | 0 | 336 | 73 | 0 | 10 | 1341 | 2218 | 30 | 83 | 14 |
| E | 301 | 1421 | 2056 | 972 | 3205 | 140 | 11085 | 326 | 128 | 2010 | 1082 | 17 | 5376 | 22 | 6155 | 6090 | 3510 | 119 | 1271 | 17 | 16483 | 3746 | 1188 | 4207 | 10 | 167 | 1341 | 1446 | 1597 |
| F | 1402 | 3 | 4 | 21 | 4 | 932 | 0 | 65 | 4 | 0 | 2 | 134 | 0 | 2 | 12 | 282 | 4 | 309 | 17 | 1 | 170 | 37 | 1 | 253 | 111 | 22 | 0 | 18 |
| G | 644 | 2 | 0 | 0 | 36 | 4310 | 0 | 0 | 46 | 0 | 84 | 342 | 223 | 0 | 159 | 112 | 21 | 218 | 1471 | 2 | 326 | 22 | 0 | 9 | 562 | 1687 | 22 | 0 | 14 |
| Ğ | 470 | 6 | 1 | 0 | 64 | 660 | 0 | 2 | 0 | 0 | 2069 | 223 | 2784 | 0 | 0 | 570 | 134 | 7 | 0 | 5 | 551 | 7 | 1 | 0 | 1499 | 182 | 2 | 10 |
| H | 4732 | 169 | 2 | 77 | 13 | 1832 | 0 | 44 | 0 | 0 | 127 | 1839 | 0 | 8 | 103 | 97 | 79 | 705 | 1 | 346 | 455 | 107 | 27 | 311 | 226 | 42 | 158 | 7 |
| I | 3 | 41 | 393 | 41 | 377 | 1 | 1215 | 2 | 1634 | 7 | 0 | 10 | 0 | 0 | 2401 | 2759 | 1998 | 9555 | 1 | 0 | 389 | 2783 | 2008 | 168 | 0 | 0 | 1 | 49 | 2518 |
| İ | 293 | 941 | 433 | 2336 | 1303 | 2011 | 454 | 221 | 1634 | 499 | 0 | 77 | 164 | 78 | 3820 | 8837 | 3150 | 14428 | 195 | 6 | 943 | 10343 | 2110 | 102 | 0 | 18 | 415 | 1706 | 4347 |
| J | 173 | 8 | 0 | 0 | 16 | 100 | 0 | 6 | 0 | 227 | 0 | 100 | 0 | 1 | 0 | 97 | 0 | 44 | 5 | 0 | 1 | 1 | 0 | 0 | 43 | 0 | 5 | 5 | 0 |
| K | 7439 | 41 | 5 | 358 | 60 | 3658 | 7 | 7 | 0 | 0 | 190 | 1718 | 3991 | 0 | 418 | 4119 | 395 | 2169 | 346 | 23 | 331 | 837 | 102 | 1965 | 1705 | 807 | 40 | 87 | 18 |
| L | 16449 | 150 | 215 | 71 | 2567 | 14020 | 18 | 929 | 0 | 0 | 29 | 4579 | 7930 | 8 | 999 | 2111 | 2949 | 705 | 6 | 29 | 7 | 377 | 0 | 685 | 2701 | 1008 | 51 | 312 | 36 |
| M | 8608 | 297 | 99 | 2 | 576 | 7844 | 6965 | 10 | 28 | 1 | 32 | 1923 | 3258 | 0 | 261 | 1076 | 103 | 597 | 705 | 9 | 38 | 289 | 23 | 20 | 1385 | 2701 | 1040 | 11 | 69 |
| N | 4141 | 220 | 1617 | 138 | 7844 | 5815 | 1 | 90 | 11147 | 1 | 48 | 5615 | 6214 | 33 | 460 | 3760 | 971 | 619 | 181 | 7 | 235 | 457 | 1158 | 43 | 1204 | 2472 | 946 | 25 | 196 |
| O | 63 | 235 | 698 | 87 | 495 | 20 | 1 | 223 | 206 | 919 | 109 | 36 | 0 | 108 | 2388 | 6638 | 3845 | 181 | 0 | 498 | 6280 | 767 | 2008 | 1209 | 264 | 117 | 572 | 171 |
| Ö | 0 | 30 | 12 | 16 | 127 | 20 | 0 | 20 | 227 | 3 | 0 | 0 | 0 | 0 | 69 | 557 | 42 | 1494 | 0 | 107 | 1186 | 167 | 39 | 130 | 280 | 0 | 61 | 595 |
| P | 2199 | 1 | 9 | 28 | 4 | 671 | 0 | 3 | 0 | 103 | 676 | 0 | 551 | 1 | 40 | 849 | 252 | 5 | 643 | 35 | 450 | 220 | 2 | 255 | 280 | 135 | 0 | 26 |
| R | 7898 | 201 | 522 | 476 | 2882 | 5415 | 98 | 441 | 0 | 0 | 109 | 5045 | 8689 | 55 | 2269 | 3236 | 1832 | 1245 | 38 | 121 | 1 | 1962 | 326 | 1619 | 2404 | 915 | 200 | 170 | 2 |
| S | 4946 | 45 | 141 | 7 | 16 | 3932 | 127 | 6 | 0 | 0 | 167 | 4173 | 5340 | 3 | 350 | 662 | 230 | 1792 | 333 | 263 | 130 | 317 | 0 | 3656 | 1369 | 698 | 84 | 425 | 70 |
| Ş | 1579 | 11 | 1 | 28 | 1 | 2068 | 0 | 135 | 11 | 0 | 29 | 1725 | 1131 | 0 | 415 | 981 | 602 | 44 | 1365 | 0 | 2 | 57 | 2 | 1162 | 312 | 356 | 190 | 64 |
| T | 6458 | 33 | 22 | 224 | 38 | 5246 | 97 | 15 | 0 | 0 | 263 | 2621 | 3990 | 5 | 268 | 1901 | 1177 | 52 | 72 | 26 | 1017 | 272 | 0 | 940 | 1558 | 1265 | 19 | 60 | 1 |
| U | 227 | 202 | 356 | 330 | 331 | 183 | 65 | 92 | 1197 | 188 | 1 | 0 | 11 | 1141 | 3061 | 1797 | 4718 | 10 | 72 | 455 | 3339 | 770 | 672 | 911 | 10 | 0 | 153 | 1531 | 15 |
| Ü | 0 | 75 | 295 | 260 | 84 | 4 | 72 | 0 | 167 | 5 | 0 | 0 | 2 | 1104 | 1166 | 873 | 3255 | 0 | 333 | 63 | 165 | 446 | 656 | 328 | 0 | 0 | 124 | 702 | 1818 |
| V | 2492 | 14 | 63 | 0 | 111 | 5475 | 0 | 31 | 0 | 13 | 26 | 1060 | 0 | 0 | 42 | 172 | 35 | 85 | 0 | 0 | 365 | 147 | 10 | 1 | 246 | 0 | 13 | 1816 |
| Y | 9093 | 93 | 50 | 0 | 527 | 4898 | 98 | 375 | 57 | 1893 | 1847 | 1 | 115 | 1971 | 136 | 4718 | 5135 | 154 | 14 | 354 | 356 | 4 | 44 | 812 | 1023 | 85 | 25 | 3 |
| Z | 1778 | 98 | 177 | 3 | 861 | 2861 | 0 | 0 | 241 | 3 | 1028 | 2265 | 1 | 13 | 1020 | 371 | 19 | 280 | 3 | 2 | 6 | 77 | 0 | 1 | 420 | 2061 | 227 | 156 |

**APPENDIX B. DISTANCES BETWEEN TWO KEYS of TWO FINGER KEYBOARD LAYOUT**

Distances between two keys of two fingers keyboard layout.

| | 1 | 2 | 3 | 4 | 5 | 6 | 7 | 8 | 9 | 10 | 11 | 12 | 13 |
|---|---|---|---|---|---|---|---|---|---|---|---|---|---|
| 1 | 0 | 1 | 2 | 3 | 4 | 1,12 | 1,8 | 2,69 | 3,2 | 4,61 | 2,24 | 2,83 | 3,61 |
| 2 | 1 | 0 | 1 | 2 | 3 | 1,12 | 1,12 | 1,8 | 2,69 | 3,2 | 2 | 2,24 | 2,83 |
| 3 | 2 | 1 | 0 | 1 | 2 | 1,8 | 1,12 | 1,12 | 1,8 | 2,69 | 2,24 | 2 | 2,24 |
| 4 | 3 | 2 | 1 | 0 | 1 | 2,69 | 1,8 | 1,12 | 1,12 | 1,8 | 2,83 | 2,24 | 2 |
| 5 | 4 | 3 | 2 | 1 | 0 | 3,2 | 2,69 | 1,8 | 1,12 | 1,12 | 3,61 | 2,83 | 2,24 |
| 6 | 1,12 | 1,12 | 1,8 | 2,69 | 3,2 | 0 | 1 | 2 | 3 | 4 | 1,12 | 1,8 | 2,69 |
| 7 | 1,8 | 1,12 | 1,12 | 1,8 | 2,69 | 1 | 0 | 1 | 2 | 3 | 1,12 | 1,12 | 1,8 |
| 8 | 2,69 | 1,8 | 1,12 | 1,12 | 1,8 | 2 | 1 | 0 | 1 | 2 | 1,8 | 1,12 | 1,12 |
| 9 | 3,2 | 2,69 | 1,8 | 1,12 | 1,12 | 3 | 2 | 1 | 0 | 1 | 2,69 | 1,8 | 1,12 |
| 10 | 4,61 | 3,2 | 2,69 | 1,8 | 1,12 | 4 | 3 | 2 | 1 | 0 | 3,2 | 2,69 | 1,8 |
| 11 | 2,24 | 2 | 2,24 | 2,83 | 3,61 | 1,12 | 1,12 | 1,8 | 2,69 | 3,2 | 0 | 1 | 2 |
| 12 | 2,83 | 2,24 | 2 | 2,24 | 2,83 | 1,8 | 1,12 | 1,12 | 1,8 | 2,69 | 1 | 0 | 1 |
| 13 | 3,61 | 2,83 | 2,24 | 2 | 2,24 | 2,69 | 1,8 | 1,12 | 1,12 | 1,8 | 2 | 1 | 0 |

# APPENDIX C. STEPS OF GENETIC ALGORITHM AND SELECTED PARAMETERS

Genetic algorithm is well-known evolutionary heuristic technique, detailed information can be found at Goldberg (1989)[1]. Pseudo-code of GA with used parameters is given below.

1. Generate randomly 100 solutions. Select the best 10 as a population.
2. **Repeat**
3. Select parents with tournament method
4. Apply two-point crossover for generate children with 0.95 crossover probability
5. Apply one point mutation with 0.01 mutation probability
6. Generate the new generation using elitist selection method
7. **Until** iteration number is two thousand

Also permutation encoding is used for the solution representation. Fitness function of GA is same as mathematical model objective function.

---

[1] Goldberg, D. E. (1989). Genetic Algorithms in Search, Optimization, and Machine Learning (1 edition). Reading, Mass: Addison-Wesley Professional.